\documentstyle[11pt,paspconf]{article}

\def\gs{\mathrel{\raise0.35ex\hbox{$\scriptstyle >$}\kern-0.6em
\lower0.40ex\hbox{{$\scriptstyle \sim$}}}}
\def\ls{\mathrel{\raise0.35ex\hbox{$\scriptstyle <$}\kern-0.6em
\lower0.40ex\hbox{{$\scriptstyle \sim$}}}}

\begin{document}

\title{Galactic morning: ALMA and the evolution of galaxies}

\author{Andrew W. Blain}
\affil{Institute of Astronomy, 
Madingley Road, 
Cambridge, CB3 0HA, UK} 

\begin{abstract}
The importance of the Atacama Large Millimeter Array (ALMA) to the study of 
high-redshift dusty gas-rich galaxies is described. ALMA will have dramatically 
greater sensitivity and angular resolution than existing millimetre(mm)/submm-wave 
telescopes. Two key areas are emphasized: i) extremely deep galaxy 
surveys down to flux densities of about 10-$\mu$Jy, two orders of 
magnitude deeper than can be reached using existing telescopes; ii) the 
study of known high-redshift galaxies, quasars and gravitational lenses. 
The sensitivity of ALMA will allow very faint galaxies to be detected, while its 
angular resolving power avoids the problem of source confusion. 
Dust continuum emission, molecular rotational line emission and atomic
fine-structure line emission will all be detected from high-redshift galaxies. 
Both the spatial structure of continuum emission and the spatial and velocity 
structure of line emission will be resolved, revealing the astrophysical 
processes at work in detail. These programs are very complementary to surveys 
using the {\it Next Generation Space Telescope (NGST)} and {\it Planck Surveyor}. 
\end{abstract}


\section{Introduction}

The determination of the intensity of extragalactic background radiation 
to within about a factor of two between the mm and ultraviolet (UV)
wavebands has been one of the most dramatic developments in observational 
cosmology over the last three years (see Fig.\,12 of Blain et al.\ 1999d). 
Others have been the routine identification of a large sample of normal 
galaxies at redshifts $z \simeq 2-4$ (Steidel et al.\ 1996), and the 
discovery of a population of very luminous dust-enshrouded star-forming 
galaxies at $z \ge 1$ using the sensitive SCUBA submm-wave camera 
at the JCMT (Holland et al.\ 1999); see the list of current data in Blain 
et al.\ (2000a). In fact, the majority ($80 \pm 50$\%) of the submm-wave 
background radiation intensity can be accounted for by the directly 
measured 850-$\mu$m flux densities of galaxies detected using SCUBA in 
the fields of gravitational lensing clusters (Smail et al.\ 1997; Blain 
et al.\ 1999b). These developments are all extremely relevant to the 
development of ALMA, which will probe the same wavelength regime. 

It is very important to make mm/submm-wave observations of 
distant galaxies. Observations of both the relative intensity of the 
submm-wave and optical background radiation, and the galaxies detected 
by SCUBA indicate that interstellar dust absorbs about 80\% of the 
optical/UV photons emitted by young stars and active galactic nuclei 
(AGNs) in the high-redshift Universe (Blain et al.\ 1999d). The absorbed energy 
warms the dust to temperatures of order 50\,K, and so it reappears as thermal 
radiation in the far-infrared (IR) waveband, and is then  
redshifted into the submm waveband. Hence, a significant fraction of 
the energy generated in galaxies and quasars may be detectable only at wavelengths 
accessible to ALMA. When combined with observations of the unobscured fraction 
of energy made using the {\it Next Generation Space Telescope} ({\it NGST}; Lilly, 
this volume), ALMA will play a key role in revealing the properties and 
structure of dust-enshrouded gas-rich galaxies at the highest redshifts. 

ALMA will have two key advantages over existing mm/submm-wave telescopes, 
including the mm-wave interferometer arrays at Nobeyama, Plateau de Bure, Owens 
Valley and Hat Creek. The collecting area of the 64 12-m ALMA antennas
will be over an order of magnitude greater, as will the 16-GHz bandwidth of the 
ALMA receivers. In addition, with baselines up to 10\,km, the resolving power 
of ALMA will be many times greater, especially at the shorter submm 
wavelengths. At present the only samples of high-redshift submm-selected galaxies 
that have been discovered using SCUBA, which has a 15-arcsec beam. It is thus 
difficult to make optical identifications and follow-up observations at other 
wavelengths; see Frayer et al.\ (1998, 1999), Hughes et al.\ (1998), 
Ivison et al.\ (1998, 2000), Smail et al.\ (1998, 1999, in prep.), 
Barger et al.\ (1999), Downes et al.\ (1999) and Lilly et al.\ (1999) 
for an indication of the difficulties involved and the successes so far. 
Because of the effects of fainter confusing sources (Blain, Ivison \& Smail 
1998), SCUBA has an effective 850-$\mu$m survey flux limit of about 2\,mJy. 
With sub-100-milliarcsec resolution, ALMA will be immune to confusion, and 
so will be able to carry out extremely deep pencil-beam surveys in all the 
atmospheric windows in the mm and submm wavebands.    

\begin{figure}[t]
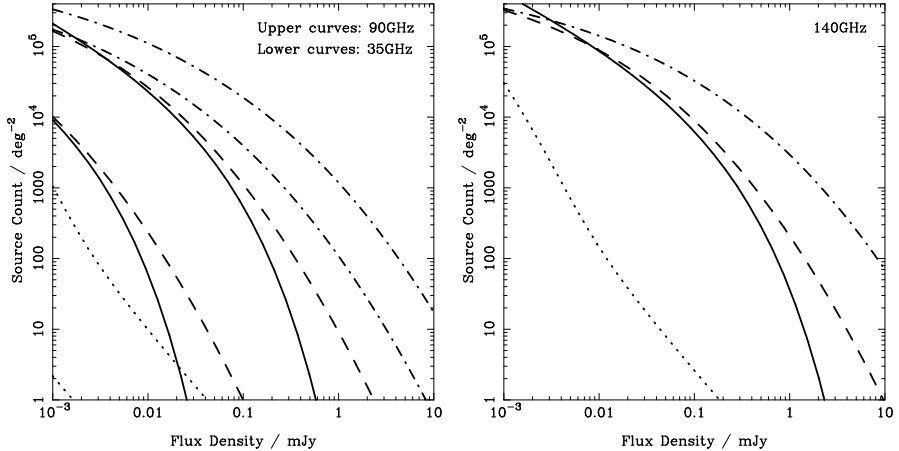

\begin{center}
\plotfiddle{blaina1a.eps}{5.1cm}{-90}{38}{38}{-230}{190}
\end{center}
\begin{center}
\plotfiddle{blaina1b.eps}{4.3cm}{-90}{38}{38}{-60}{337}
\end{center}
\vskip -5cm
\caption{Counts of continuum and line emitting galaxies expected at flux 
densities accessible to ALMA in the 35-, 90- and 140-GHz bands. The continuum 
and line counts are related by assuming a linewidth 
of 300\,km\,s$^{-1}$. The counts of continuum objects expected in the MG model 
and from merging galaxies in the HC model are represented by solid and dashed 
lines respectively. The counts of quiescent non-merging galaxies in the HC 
model are represented by dotted lines. The counts of line-emitting 
galaxies derived from the MG model (see Blain et al.\ 2000b) are represented by 
dot-dashed lines.}
\end{figure}

\begin{figure}[t]
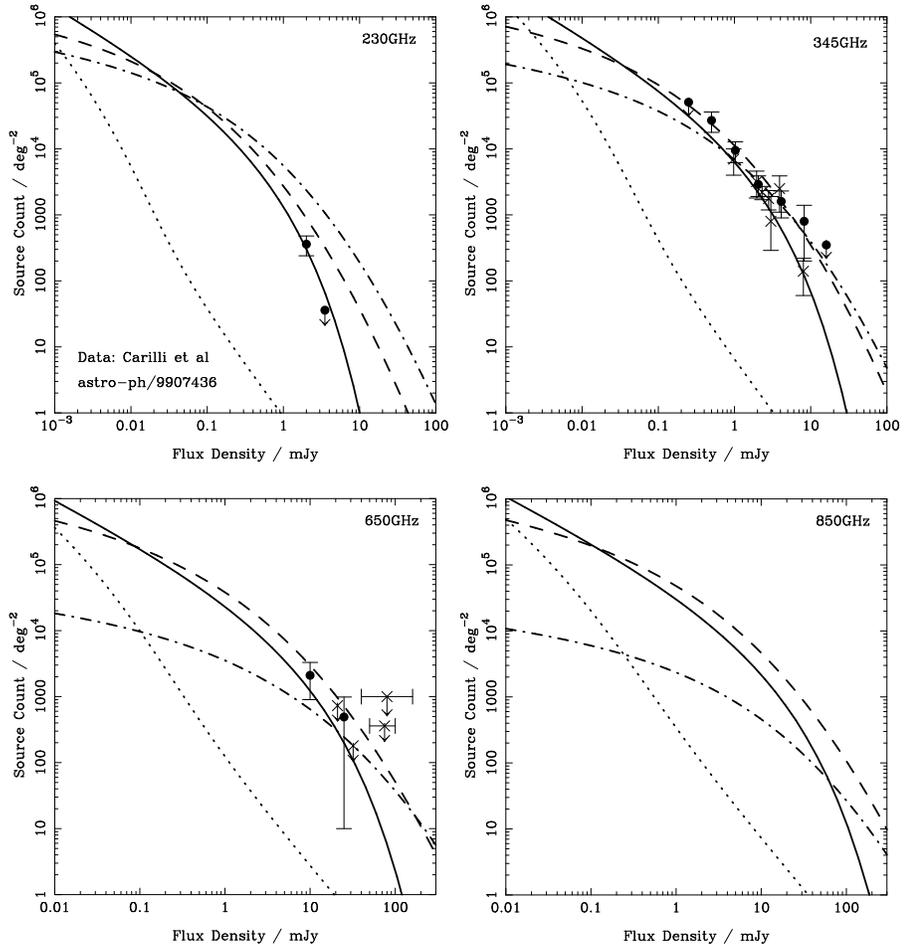

\begin{center}
\plotfiddle{blaina2a.eps}{5.1cm}{-90}{38}{38}{-230}{190}
\end{center}
\begin{center}
\plotfiddle{blaina2b.eps}{4.3cm}{-90}{38}{38}{-60}{337}
\end{center}
\begin{center}
\plotfiddle{blaina2c.eps}{5.1cm}{-90}{38}{38}{-230}{325}
\end{center}
\begin{center}
\plotfiddle{blaina2d.eps}{4.3cm}{-90}{38}{38}{-60}{472}
\end{center}
\vskip -9.8cm 
\caption{Counts of continuum and line emitting galaxies expected 
in the 230-, 345-, 650- and 850-GHz bands (line styles as in Fig.\,1). The 
345- and 650-GHz data is described in Blain et al.\ (2000a).} 
\end{figure} 

\section{The population of high-redshift galaxies for ALMA to observe} 

Based on the background radiation spectrum determined using the FIRAS and DIRBE  
instruments on the {\it COBE} satellite (Puget et al.\ 1996; Fixsen et al.\ 1998; 
Hauser et al.\ 1998; Schlegel, Finkbeiner \& Davis 1998), the surface densities of 
galaxies detected using SCUBA at 850-$\mu$m (see Blain et al.\ 2000a), the IRAM 30-m 
telescope at 1.25\,mm (Carilli et al.\ 2000) and {\it ISO} at 175\,$\mu$m (Kawara 
et al.\ 1998; Puget et al.\ 1999), it is possible to construct a consistent 
picture of the evolution of dusty galaxies that are very luminous in the
far-IR and submm wavebands (Blain et al.\ 1999c,d). The models developed can be 
used to extrapolate to wavelengths and flux densities that are not accessible to 
current instruments, but are relevant to making predictions for ALMA. Observations 
of CO rotational line emission from two galaxies detected using SCUBA 
(Frayer et al.\ 1998, 1999), have allowed these predictions to be extended to 
include observations of molecular and atomic fine structure lines 
(Blain et al.\ 2000b). The resulting counts $N(>S)$ are shown in Figs\,1 \& 2; 
two models for the population of very luminous dusty galaxies are assumed. First, 
the modified Gaussian (MG) model (Blain et al.\ 1999c; Barger et al.\ 1999), in 
which the low-redshift luminosity function of {\it IRAS} galaxies undergoes 
pure luminosity evolution. Secondly, an hierarchical (HC) model (Blain et al.\ 
1999d), in which luminous galaxies, with a dust temperature assumed to be 40\,K, 
are associated with the mergers of dark matter halos. The counts 
of non-merging, quiescent galaxies expected between merger events in the HC 
model are also shown. A cooler dust temperature of 20\,K, typical of low-redshift 
spiral galaxies, is assumed for the quiescent galaxies. The evolution of their 
specific star-formation rate is assumed to trace that of the merging galaxies, 
and at $z=0$ they are assumed to be 100 times less luminous than merging 
galaxies with the same mass. The counts of quiescent galaxies are only expected to  
exceed those of merging galaxies at the faintest flux densities. 

As more data becomes available over the next few years, from {\it SIRTF} and 
BOLOCAM, and in particular from complete redshift distributions of the galaxies 
detected using SCUBA and {\it ISO}, these predictions will be refined; 
however, they already allow the key scientific strengths of ALMA to be explained. 

\section{ALMA sensitivities and detection rates} 

Based on information at the ALMA website (www.nrao.edu/alma), the crucial details 
required to assess the best strategies for ALMA surveys are listed in Table\,1. 
The key parameters are the sensitivity $S$ -- the depth reached 
by the array in a single pointing in a 1-s integration -- and the primary beam area 
$A_{\rm p}$ -- the area of the field that is observed in a single pointing. In order to 
assess the performance of ALMA in each of its seven bands, the time $t$ required to 
survey an area $A$ to a flux density $S_L$ can be estimated; 
$t \propto (S_L/S)^{-1/2} (A/A_{\rm p})$. If $S_L$ is fixed, then $A(t)$ can be 
determined, and so the number of sources detected in time $t$, the product of 
$A(t)$ and the count of galaxies $N(>S_L)$ (Figs\,1 \& 2), can be predicted. 
Note that the detection rate depends on both the performance of ALMA and the 
properties of the population of IR-luminous galaxies. The results 
of this procedure, which is outlined in more detail by Blain \& Longair (1996) and 
Blain (1999), are shown in Fig.\,3, for both continuum and line observations. 

The detection rate for continuum sources (Fig.\,3; left) is expected to be 
greatest in the 230- and 345-GHz bands. For line emission, the greatest 
detection rates are expected at 140 and 230\,GHz (Fig\,3; right). The 
differences in the detection rates in the different bands are due to a 
trade-off between the increase in the counts of galaxies and the 
decrease in the sensitivity and primary beam area as the observing frequency 
increases. The detection rates are expected to be greatest for continuum flux 
densities between 1 and 10\,mJy. This flux density range can be observed to a 
significant signal-to-noise ratio in a single ALMA primary beam in only a few 
seconds at 230\,GHz, and can also be reached using existing submm-wave 
telescopes. As shown in Fig.\,3 (left), the detection rate of continuum sources 
expected using future wide-field imaging instruments on single-antenna 
telescopes (for example 
BOLOCAM and SCUBA-2) at these flux densities will be comparable to that using ALMA. 

\begin{table}[t]
\caption{Sensitivities in continuum ($S_{\rm c}$) and line ($S_{\rm l}$) observations, 
each for making a 10-$\sigma$ detection in 1\,s in each of the ALMA observing 
bands at frequency $\nu$. For spectral lines, a velocity resolution 
$\Delta v = 300$\,km\,s$^{-1}$ is assumed. The sensitivities integrate 
down with integration time $t$ as $t^{-1/2}$. The spectral line sensitivity 
depends on the velocity resolution as $\Delta v^{-1/2}$. The approximate 
full-width-half-maximum primary 
beamwidth $\theta_{\rm p}$ and beam area $A_{\rm p}$ are also listed.} 
\begin{center}\scriptsize
\begin{tabular}{lcccc}
$\nu$ / GHz & $\theta_{\rm p}$ / arcsec & $A_{\rm p}$ / arcmin$^{2}$ & 
$S_{\rm c}$ / mJy & $S_{\rm l}$ / mJy \\  
\tableline
35 & 175 & 6.8 & 1.16 & 17.2 \\
90 & 69 & 1.02 & 2.40 & 22.3 \\
140 & 44 & 0.42 & 2.71 & 20.1 \\
230 & 27 & 0.16 & 4.65 & 26.9 \\
345 & 18 & 0.069 & 10.8 & 49.3 \\
650 & 9.5 & 0.020 & 147 & 515 \\
850 & 7.2 & 0.011 & 271 & 828 \\
\end{tabular}
\end{center}
\end{table}

The unique angular resolving power and sensitivity 
of ALMA would be better used for making deeper surveys over smaller regions of 
the sky, detecting sources that are too faint for existing interferometers, and 
which are much fainter than the confusion limit for single-antenna telescopes. 
Although the detection rate in a deeper ALMA survey will be lower, a 
230-GHz survey down to a flux density of 20\,$\mu$Jy (100 times deeper than 
that currently possible) is still expected to detect a galaxy every 5\,hours. At 
this depth, ALMA is still far from the confusion limit, and so an even deeper survey 
would be possible, if the initial results showed that it would be scientifically 
interesting. The same is generally true for a survey targeted at the detection of line 
emission -- the greatest detection rates are expected at flux densities that do 
not fully exploit the unique capabilities of ALMA, and so ultradeep surveys 
should be the goal. 

\begin{figure}[t]
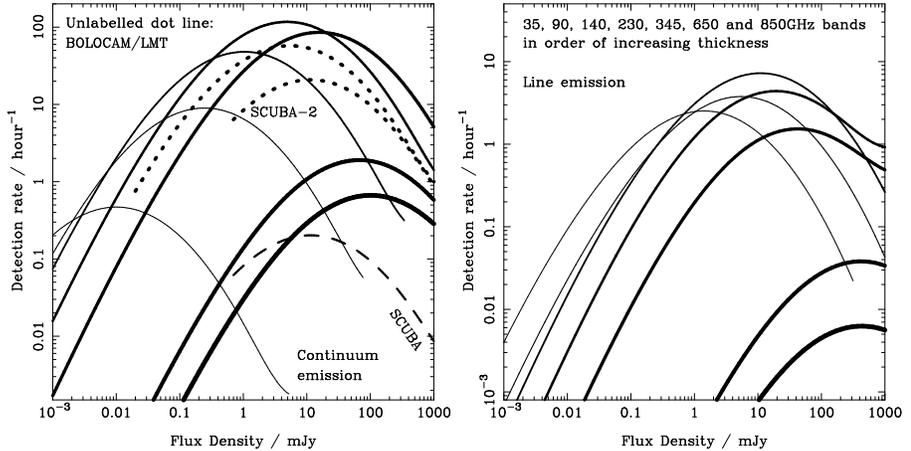

\vskip -5cm  
\begin{center}
\plotfiddle{blaina3a.eps}{5.1cm}{-90}{38}{38}{-230}{190}
\end{center}
\begin{center}
\plotfiddle{blaina3b.eps}{4.3cm}{-90}{38}{38}{-60}{337}
\end{center}
\vskip -5cm
\caption{Continuum (left; 5-$\sigma$) and line (right; 10-$\sigma$) detection 
rates expected in surveys in all seven ALMA observing bands. 
The results are obtained by combining the sensitivities listed in Table\,1 
with the MG model counts shown in Figs\,1 \& 2. 
For comparison, the performance of the JCMT's SCUBA camera at 850\,$\mu$m and 
the expected performance of the 1.1-mm BOLOCAM camera (Glenn et al.\ 1998) and 
the proposed 850-$\mu$m SCUBA-2 camera (Holland 1999) above their 
respective confusion limits are also shown -- ALMA will not be affected. 
The line predictions assume a 300-km-s$^{-1}$ line and a 16-GHz band.
}
\end{figure}

\subsection{Combining line and continuum observations} 

Existing mm/submm-wave interferometers observe spectra over 
a bandwidth $\Delta \nu$ of only about 1\,GHz simultaneously. This is an important 
limitation for two related reasons. First, in order to detect the line emission 
from a high-redshift galaxy in a transition listed in Table\,2 using such an 
instrument, the redshift of the 
emitting source must be known to better than about 1\%, because 
$\Delta z = \Delta \nu / \nu_{\rm line}$. Currently, optical/near-IR 
spectroscopy with a reasonable signal-to-noise ratio (Ivison et al.\ 1998; 
Barger et al.\ 1999) is required before mm-wave lines can be
observed in submm-selected galaxies (Frayer et al.\ 1998, 1999). This is  
often a difficult proposition. Secondly, in a blank-field observation, each line
is currently visible in only a small redshift interval, and so unbiased searches 
for high-redshift line-emitted galaxies are not yet possible (Blain et al.\ 2000b). 
The 16-GHz bandwidth of ALMA will make these observations much easier. 

The observed separation of adjacent CO lines from a galaxy at redshift $z$ is 
115\,GHz$/(1+z)$. This is 
greater than 16\,GHz for reasonable redshifts $z \le 6$, and so ALMA is typically 
unlikely to detect two adjacent CO lines from the same object without retuning
the receivers. For $z \simeq 3$, the separation of adjacent lines is 28.8\,GHz, 
and so in two deep integrations with adjacent tunings, say 210-226 and 
226-242\,GHz, ALMA should detect pairs of lines from at least half of the $z=3$ 
galaxies, and so provide a direct measurement of their redshifts. 

In a deep single-tuned observation, CO line emission from a subset of the 
galaxies detected in the continuum, and a number of faint lines for which there is 
no detected continuum emission, should both be detected. 
As only one line from the ladder of CO transitions is expected per source, an 
unequivocal redshift cannot be derived directly. For example, a line detected at 
220\,GHz could be CO(2$\rightarrow$1) at $z=0.045$, (3$\rightarrow$2) at 
$z=0.57$, (4$\rightarrow$3) at $z=1.10$, or (5$\rightarrow$4) at $z=1.62$ etc. 
Retuned observations, targeted at a different transition would discriminate between 
these cases, but in general this would be a time-consuming exercise. An alternative, 
in at least a statistical sense, would be to use the line:continuum ratio, which  
can be expressed as an equivalent width in frequency, to estimate the 
$J$ value of the line, and thus the redshift. The line:continuum ratio is expected 
to decrease as the value of $J$ increases; see Table\,2. Equivalent widths of 8.8, 
4.7, 2.9 and 1.8\,GHz are expected for the four lines in the example above, and 
so if both the line and continuum flux can be measured to an accuracy of 10\%, 
then there are reasonable prospects of discriminating between CO transitions using 
this method. A measurement of the slope of any detected continuum emission 
across the band, and of the continuum colors from band-to-band could also be 
used to provide a little extra redshift information, as the thermal continuum 
spectrum will begin to flatten into the restframe far-IR waveband. The 
reliability of these techniques could be verified easily using conventional 
multi-line observations. 

\begin{table}[t]
\caption{Frequencies $\nu_{\rm line}$ and $z=0$ equivalent widths (EWs) of 
CO molecular line emission expected in the model 
used to predict line counts in this paper (Blain et al.\ 2000b). 
The ratio of line to continuum power decreases steadily as line 
excitation increases. The EWs decrease with redshift as $(1+z)^{-1}$. 
This feature offers hope for the use of the detection of a single 
line to provide a preliminary redshift for a galaxy.}
\begin{center}\scriptsize
\begin{tabular}{cccccc}
CO transition & $\nu_{\rm line}$/GHz & EW / GHz &
CO transition & $\nu_{\rm line}$/GHz & EW / GHz \\
\tableline
1$\rightarrow$0 & 115 & 13.4 & 5$\rightarrow$4 & 576 & 4.7\\
2$\rightarrow$1 & 230 & 9.2 & 6$\rightarrow$5 & 691 & 3.3\\
3$\rightarrow$2 & 346 & 7.4 & 7$\rightarrow$6 & 807 & 1.8\\
4$\rightarrow$3 & 461 & 6.1 & 8$\rightarrow$7 & 922 & 0.4\\
\end{tabular}
\end{center}
\end{table}

\section{Practical surveys}

ALMA is uniquely well equipped to carry out two distinct kinds of survey; surveys 
to depths that are impossible using other facilities because of source confusion, 
and rapid systematic high-resolution observations of high-redshift galaxies 
that have been detected elsewhere. Because of its great sensitivity, 
integration times per source will be short, and many will be spatially resolved.  

\subsection{Ultradeep mm/submm-wave surveys} 

ALMA is the only submm-wave telescope with the resolving power and 
collecting area to carry out sensitive continuum surveys at flux densities 
below 0.1\,mJy, flux densities at which $L^*$ galaxies would be detectable at 
moderate redshifts. The results will discriminate between currently acceptable 
models of galaxy evolution; see Fig.\,2. Because of the large 16-GHz bandwidth 
of ALMA, spectral line surveys will be comparable in efficiency to those in 
the continuum at the most favored observing frequencies, and will be more efficient 
in the observing bands at lower frequencies. The expected detection rates of 
both line and continuum sources are shown in Fig.\,3. 

\subsection{Follow-up mm/submm-wave surveys} 

ALMA will be a very powerful instrument for making follow-up observations of 
sources that are detected using other instruments, in the X-ray, optical, 
near-IR, mid-IR, far-IR and mm/submm wavebands. 

For example, optically-selected Lyman-break galaxies are typically expected to 
have 850-$\mu$m flux densities of order 0.2\,mJy (Chapman et al.\ 2000; 
Peacock et al.\ 2000). A 5-$\sigma$ detection of these galaxies at this 
frequency could be made in about 10\,min using ALMA. Similar investigations to 
search for dust continuum and spectral line emission from large numbers of 
optical-, IR- and X-ray-selected high-redshift QSOs, in 
radio galaxies and even in gamma-ray bursts (Blain \& Natarajan 2000) can also be 
conducted. 

Galaxies discovered at flux densities $\gs 100$\,mJy at 175\,$\mu$m using 
{\it ISO} (Kawara et al.\ 1998; Puget et al.\ 1999) would be expected to 
have flux densities of order 5\,mJy at 850\,$\mu$m if $z \simeq 1$. Much larger 
samples of similar galaxies are expected to be compiled during the next few years 
using {\it SIRTF} and {\it IRIS/ASTRO-F}. The small aperture of these facilities means 
that the positional uncertainty of the detected sources is large -- about 1\,arcmin. 
However, ALMA is sufficiently sensitive to make a rapid scan over a 15-beam, 
1-arcmin$^2$ field to this depth in less than 1\,min to locate the emission. 
A high signal-to-noise ALMA image of the located source could then be made in less 
than 10\,s in each observing band. 

Large samples of submm-selected galaxies will be produced by the BOLOCAM and 
SCUBA-2 cameras on the ground and by the {\it Planck Surveyor}, {\it H2L2} and 
{\it FIRST} space missions. ALMA and {\it Planck} are exceedingly complementary 
facilities. While the {\it Planck} all-sky submm-wave survey is primarily intended 
to map structure in the cosmic microwave background (CMB) radiation in unprecedented 
detail at angular scales greater than 4\,arcmin, of order $10^5$ galaxies should 
also be detected at a frequency of 350\,GHz (Blain 1998). Each galaxy will have 
a flux density greater than 100\,mJy. At the same frequency, ALMA can scan a 
16-arcmin$^2$ region to the 10-$\sigma$ sensitivity limit of 100\,mJy that is 
required to locate the position of the {\it Planck} source to within 1\,arcsec in 
only 3\,s. 
This observation would require only 12\,ms of observing time in each primary beam, 
if efficient mosaicking of different fields can be achieved. The required scan 
rate of up to 0.5\,deg\,s$^{-1}$ is within the specifications of the ALMA 
antennas. Hence, ALMA could locate {\it Planck} sources very efficiently, and 
then observe them at all available wavelengths in only a few tens of seconds 
per object. The result will be a catalog of all the most luminous submm 
galaxies and AGN on the sky, a significant fraction (up to 10\%) of which are 
expected to be gravitational lenses (Blain 1996, 1998). The results will also 
reduce the level of foreground contamination of the {\it Planck} CMB signal.
 
\section{Gravitational lenses} 

\subsection{Lensing by clusters} 

Gravitational lensing by clusters of galaxies was exploited in the first 
submm-wave surveys to increase the effective sensitivity of the 
SCUBA instrument and to reduce the effects of source confusion (Smail et al.\ 1997). 
ALMA will have such high resolution that it can carry out critical line mapping -- 
targeted imaging of high-magnification regions of the background sky behind 
rich clusters of galaxies. Magnification 
factors of several tens are known to be found in these regions for high-redshift 
galaxies, and so galaxies with unmagnified flux densities less than 
1\,$\mu$Jy could plausibly be detected. These flux densities correspond to galaxies 
with luminosities $\simeq 10^9$\,L$_\odot$, much less than $L^*$, at very high 
redshifts. The large magnification in these regions of clusters are associated 
with strong shearing of the images, and so the submm emission from dust and gas in 
the star-forming 
constituent fragments of primordial galaxies could be resolved by ALMA, 
perhaps allowing their morphologies and internal dynamics to be reconstructed. 

\subsection{Lensing by galaxies} 

The detailed, resolved CO maps made of known gravitational lenses, for 
example the Cloverleaf QSO (Kneib et al.\ 1998), have demonstrated directly that 
existing mm-wave interferometers can be used to image such objects. 
The much greater resolving power and sensitivity of ALMA will allow these 
observations to be carried out routinely in only a few minutes. 
Observations of fainter lensed galaxies using ALMA, and direct searches 
for lensed background galaxies in the fields of elliptical galaxies and 
edge-on spiral galaxies (Blain, M\"oller \& Maller 1999a) 
should allow both extremely faint counts of dusty galaxies to be determined 
and the mass distribution of dark-matter halos to be probed. 

\subsection{The lens/foreground galaxies} 

Molecular absorption lines in galaxies along the line of sight to 
luminous continuum radiation sources have been studied (see for example
Wiklind \& Combes 1996). These observations reveal details of the 
interstellar medium (ISM) of normal galaxies at moderate redshifts. In the case of 
strong gravitational lenses, several different lines of sight through the 
lensing galaxy, provided by the multiple images, can be studied. The results may 
allow differential extinctions to be measured, and thus provide more information 
with which to constrain both the mass distribution within the lens and the 
geometry and expansion rate of the Universe, in addition to providing templates for
extinction curves in high-redshift galaxies. This type of 
observation will be very easy using ALMA. The internal dynamics of cold gas in 
lensing galaxies will be resolved in detail and accurate redshifts found.

\section{Conclusions} 

ALMA will be a submm instrument of unprecedented sensitivity and resolving 
power, both for continuum and line observations. Ultradeep surveys for dust 
and gas in high-redshift galaxies will allow the progress of galaxy evolution 
to be traced in more detail, both as a general process and from galaxy to galaxy. 
High-resolution, high signal-to-noise follow-up observations of galaxies 
detected using other instruments will allow the physical conditions in the 
ISM of a wide range of high-redshift galaxies to be probed. ALMA will have 
the sensitivity, resolving power, and wide spectroscopic bandwidth to search for 
and to study galaxies that are gravitationally lensed by galaxies, groups and 
clusters, allowing its performance to be enhanced further by exploiting 
gravitational telescopes. 

\acknowledgments

The author, Raymond and Beverly Sackler Foundation Research Fellow, 
gratefully acknowledges support from the Foundation within the IoA 
Deep Sky Initiative, and from the `Lensnet' EC TMR network. Thanks 
to Ian Smail, Rob Ivison, Jean-Paul Kneib, Dave Frayer, 
Jaime Bock, Nick Scoville, Ole M\"oller, Allon Jameson, Malcolm Longair, Amy Barger, 
Len Cowie and Priya Natarajan and Kate Quirk for collaborative work and 
comments on the manuscript.

\end{document}